\documentclass{article}

\input{tcilatex}

\begin{document}

\begin{center}
\textbf{AN ALGORITHM FOR GENERATING ROTATING BRANS-DICKE WORMHOLE SOLUTIONS}

\textbf{\bigskip }

Kamal K. Nandi$^{1,3,a}$ and Yuan-Zhong Zhang$^{2,3,b}$

$^{1}$Department of Mathematics, University of North Bengal, Darjeeling
(W.B.) 734 430, India

$^{2}$CCAST (World Laboratory), Chinese Academy of Sciences, P.O. Box 8730,
Beijing 100080, China

$^{3}$Institute of Theoretical Physics, Chinese Academy of Sciences, PO Box
2735, Beijing 100080, China

$^{a}$Email: kamalnandi1952@yahoo.co.in

$^{b}$Email: yzhang@itp.ac.cn

\bigskip

\textbf{Abstract}
\end{center}

\textit{Using the ansatz of Matos and N\'{u}\~{n}ez, the present
article proposes an algorithm for generating several classes, not
all independent, of asymptotically flat rotating wormhole
solutions in the Brans-Dicke Theory. The algorithm allows us to
associate a real number }$n$ \textit{with} \textit{each static and
generated rotating solution. We shall also demonstrate how to
match a rotating wormhole to a flat spacetime at the matter
boundaries. The vacuum string extensions of the solutions are
straightforward. The physical interpretations of the solutions are
deferred.}

\begin{center}
\bigskip

\textbf{I. Introduction}
\end{center}

Recently, there is a revival of interest in the Brans-Dicke Theory (BDT) due
principally to the following reasons: The theory occurs naturally in the low
energy limit of the effective string theory in four dimensions or the
Kaluza-Klein theory. It is found to be consistent not only with the weak
field solar system tests but also with the recent cosmological observations.
Moreover, the theory accommodates Mach's principle. (It is known that
Einstein's General Relativity (EGR) can not accommodate Mach's principle
satisfactorily). All these are well known.

A less well known yet an important arena where BDT has found immense
applications is the field of wormhole physics, a field recently re-activated
by the seminal work of Morris, Thorne and Yurtsever (MTY) [1]. Conceptual
predecessors of MTY wormholes could be traced to the geometry of Flamm
paraboloid, Wheeler's concept of \textquotedblleft charge without
charge\textquotedblright , Klein bottle or the Einstein-Rosen bridge model
of a particle [2]. Wormholes are topological handles that connect two
distant regions of space. These objects are invoked in the investigations of
problems ranging from local to cosmological scales, not to mention the
possibility of using these objects as a means of interstellar travel [1].
Wormholes require for their construction what is called \textquotedblleft
exotic matter\textquotedblright\ - matter that violate some or all of the
known energy conditions, the weakest being the averaged null energy
condition. Such matters are known to arise in quantum effects. However, the
strongest theoretical justification for the existence of exotic matter comes
from the notion of dark energy or phantom energy that are necessary to
explain the present acceleration of the Universe. Some classical fields can
be conceived to play the role of exotic matter. They are known to occur, for
instance, in the $R+R^{2}$ theories [3], Visser's thin shell geometries [4]
and, of course, in scalar-tensor theories [5] of which BDT is a prototype.
There are several other situations where the energy conditions could be
violated [6].

BDT describes gravitation through a metric tensor ( $g_{\mu \nu }$) and a
massless scalar field ( $\phi $). The BD action for the coupling parameter $%
\omega =-1$ can be obtained in the Jordan frame from the vacuum linear
string theory in the low energy limit. The action can be conformally
rescaled into what is known as the Einstein frame action in which the scalar
field couples minimally to gravity. The last is referred to as the Einstein
minimally coupled scalar field theory (EMS). Several static wormhole
solutions in EMS and BDT have been widely investigated in the literature
[7]. However, exact rotating wormhole solutions are relatively scarce,
especially, in the BDT except a recent one in EMS discussed by Matos and N%
\'{u}\~{n}ez [8]. In this context, we recall the well known fact that the
formal independent solutions of BDT are \textit{not} unique. (Of course, the
black hole solution \textit{is} unique for which the BD/EMS scalar field is
trivial in virtue of the so called \textquotedblleft no scalar hair"
theorem.) Four classes of static BDT solutions were derived by Brans [9]
himself way back in 1962, and the corresponding four classes of EMS
solutions are also known [10]. But recently it has been shown that only two
of the four classes of Brans' solutions are independent [11]; the other two
can be derived from them. However, the forms of all the original four
classes of Brans' or EMS solutions are suggestive in their own right and we
shall consider all of them as seed solutions.

In this article, we shall derive three classes of asymptotically flat
rotating wormholes in the EMS and BDT. The remaining class of solutions
(class III) is not asymptotically flat and hence will not be discussed here.
Our strategy is to start from the static EMS solutions and then generate
rotating solutions in the EMS since they involve less number of identified
constants than in the BDT. We shall then transfer them back into those of
the BDT. The BDT solutions can further be rephrased as solutions of the
vacuum 4-dimensional low energy string theory ( $\omega =-1$). Throughout
the article, we take units such that $8\pi G=c=1$.

\begin{center}
\textbf{II. The action, ansatz and the algorithm}
\end{center}

Let us start from the 4-dimensional, low energy effective action of
heterotic string theory compactified on a 6-torus. The tree level string
action, keeping only linear terms in the string tension $\alpha ^{\prime }$
and in the curvature $\widetilde{R}$, takes the following form in the matter
free region ( $S_{matter}=0$):%
\begin{equation}
S_{string}=\frac{1}{\alpha ^{\prime }}\int d^{4}x\sqrt{-\widetilde{g}}e^{-2%
\widetilde{\Phi }}\left[ \widetilde{R}+4\widetilde{g}^{\mu \nu }\widetilde{%
\Phi }_{,\mu }\widetilde{\Phi }_{,\nu }\right] ,
\end{equation}%
where $\widetilde{g}^{\mu \nu }$ is the string metric and $\widetilde{\Phi }$
is the dilaton field. Note that the zero values of other matter fields do
not impose any additional constraints either on the metric or on the
dilation [12]. Under the substitution $e^{-2\widetilde{\Phi }}$ $=\phi $,
the above action reduces to the BD action%
\begin{equation}
S_{BD}=\int d^{4}x\sqrt{-\widetilde{g}}\left[ \phi \widetilde{R}+\frac{1}{%
\phi }\widetilde{g}^{\mu \nu }\phi _{,\mu }\phi _{,\nu }\right] ,
\end{equation}%
in which the BD coupling parameter $\omega =-1$. This particular value is
actually model independent and it actually arises due to the target space
duality. It should be noted that the BD action has a conformal invariance
characterized by a constant gauge parameter $\xi $ [13]. Arbitrary values of
can actually lead to a shift from the value $\omega =-1$, but we fix this
ambiguity by choosing $\xi =0$. Under a further substitution%
\begin{eqnarray}
g_{\mu \nu } &=&\phi \widetilde{g}_{\mu \nu }  \nonumber \\
d\varphi &=&\sqrt{\frac{2\omega +3}{2\alpha }}\frac{d\phi }{\phi };\alpha
\neq 0;\omega \neq \frac{3}{2},
\end{eqnarray}%
in which we have introduced, on purpose, a constant parameter $\alpha $ that
can have any sign. Then the action (2) goes into the form of EMS action%
\begin{equation}
S_{EMS}=\int d^{4}x\sqrt{-g}\left[ R+\alpha g^{\mu \nu }\varphi _{,\mu
}\varphi _{,\nu }\right] .
\end{equation}%
The EMS field equations are given by%
\begin{eqnarray}
R_{\mu \nu } &=&-\alpha \varphi _{,\mu }\varphi _{,\nu }  \nonumber \\
\varphi _{;\mu }^{:\mu } &=&0
\end{eqnarray}%
We shall choose $\alpha =+1$, $\varphi =\varphi (l)$ in what follows, and
adopt the Matos-N\'{u}\~{n}ez ansatz [8]:%
\begin{equation}
ds^{2}=-f(l)\left( dt+a\cos \psi d\psi \right) ^{2}+\frac{1}{f(l)}\left[
dl^{2}+(l^{2}+l_{0}^{2})\left( d\theta ^{2}+\sin ^{2}\theta d\psi
^{2}\right) \right] ,
\end{equation}%
where $l_{0}$ is an arbitrary constant, $a$ is identified as the rotational
parameter and $f(l)$ is a solution of the field equations%
\begin{eqnarray}
\left[ \left( l^{2}+l_{0}^{2}\right) \frac{f^{^{\prime }}}{f}\right]
^{\prime }+\frac{a^{2}f^{2}}{l^{2}+l_{0}^{2}} &=&0, \\
\left( \frac{f^{^{\prime }}}{f}\right) ^{2}+\frac{4l_{0}^{2}+a^{2}f^{2}}{%
\left( l^{2}+l_{0}^{2}\right) ^{2}}-2\varphi ^{\prime 2} &=&0,
\end{eqnarray}%
where the prime denotes differentiation with respect to $l$.

\textit{Algorithm:}

Let $f_{0}\equiv f(l;p,q,a=0)$ be a given solution of the static
configuration in which $p,$ $q$ are arbitrary constants in the solution.
(Combinations of these constants can be interpreted as the mass and scalar
charge of the configuration.) Then the rotating solution is
\begin{equation}
f(l;p,q,a)=\frac{2npq\delta f_{0}^{-1}}{a^{2}+n\delta ^{2}f_{0}^{-2}}
\end{equation}%
where $n$ is a real number specific to a given static solution and $\delta $
is a free parameter allowed by the rotating solution. The scalar field $%
\varphi $ is remarkably given by the static solution of the massless
Klein-Gordon equation $\varphi _{;\mu }^{:\mu }=0$. The static solution ( $%
a=0$) following from Eq.(9) gives $\delta =2pq$. For the rotating solution,
the value of $\delta $ may be fixed either by the condition of asymptotic
flatness or via the matching conditions at specified boundaries. \textit{%
Eq.(9) is the algorithm we propose.} Matos and N\'{u}\~{n}ez [8] defined the
free parameter as $\delta =\sqrt{D}$ . The difficulty is that the field
Eqs.(7) and (8) then identically fix $\delta ^{2}=D=0$ giving $f=0$ which
obviously yields a meaningless solution when put in the metric (6). When we
put Eq.(9) in Eqs. (7) and (8), we find that they reduce to differential
equations for $a=0$ and $f=f_{0}$. This fact completely justifies our
algorithm.

\begin{center}
\textbf{III. BD Class I rotating solution }
\end{center}

Let us now consider the Class I EMS solution due to Buchdahl [14]:%
\begin{eqnarray}
ds^{2} &=&-\left( \frac{1-\frac{m}{2r}}{1+\frac{m}{2r}}\right) ^{2\beta
}dt^{2}+\left( 1-\frac{m}{2r}\right) ^{2(1-\beta )}\left( 1+\frac{m}{2r}%
\right) ^{2(1+\beta )}\times  \nonumber \\
&&\lbrack dr^{2}+r^{2}d\theta ^{2}+r^{2}\sin ^{2}\theta d\psi ^{2}]
\end{eqnarray}%
\begin{equation}
\varphi (r)=\sqrt{\frac{2(\beta ^{2}-1)}{\alpha }}\ln \left[ \frac{1-\frac{m%
}{2r}}{1+\frac{m}{2r}}\right] ,
\end{equation}%
where $m$ and $\beta $ are two arbitrary constants. The metric (6) can be
expanded to give%
\begin{eqnarray}
ds^{2} &=&-\left[ 1-\frac{2M}{r}+\frac{2M^{2}}{r^{2}}+O\left( \frac{1}{r^{3}}%
\right) \right] dt^{2}+\left[ 1+\frac{2M}{r}+O\left( \frac{1}{r^{3}}\right) %
\right] \times  \nonumber \\
&&\lbrack dr^{2}+r^{2}d\theta ^{2}+r^{2}\sin ^{2}\theta d\psi ^{2}],
\end{eqnarray}%
from which one can read off the Keplerian mass%
\begin{equation}
M=m\beta
\end{equation}%
The metric has a naked singularity at $r=m/2$. For $\beta =1$, it reduces to
the Schwarzschild solution in isotropic coordinates. For $\alpha =+1$ and $%
\beta >1$, it represents a traversable wormhole. It is symmetric under
inversion of the radial coordinate $r\rightarrow \frac{1}{r}$ and we have
two asymptotically flat regions (at $r=0$ and $r=\infty $) connected by the
throat occurring at $r_{0}^{+}=\frac{m}{2}\left[ \beta +\sqrt{\beta ^{2}-1}%
\right] $. Thus real throat is guaranteed by $\beta ^{2}>1$. For the choice $%
\alpha =+1$, the quantity $\sqrt{2(\beta ^{2}-1)}$ is real such that there
is a real scalar charge $\sigma $ from Eq.(10) given by%
\begin{equation}
\varphi =\frac{\sigma }{r}=-\frac{2m}{r}\sqrt{\frac{\beta ^{2}-1}{2}}
\end{equation}%
But, in this case, we have violated almost all energy conditions in
importing \textit{by hand }a negative sign before the kinetic term in
Eq.(5). Alternatively, we could have chosen $\alpha =-1$ in Eq.(10), giving
an imaginary charge $\sigma $. In both cases, however, we end up with the
same equation $R_{\mu \nu }=-\varphi _{,\mu }\varphi _{,\nu }$. There is
absolutely no problem in accommodating an imaginary scalar charge in the
wormhole configuration [15,16].

Using the transformation $l=r+\frac{m^{2}}{4r}$, the solution (10) and (11)
can be expressed as%
\begin{eqnarray}
ds^{2} &=&-f_{0}(l)dt^{2}+\frac{1}{f_{0}(l)}\left[ dl^{2}+(l^{2}-m^{2})%
\left( d\theta ^{2}+\sin ^{2}\theta d\psi ^{2}\right) \right] ,  \nonumber \\
f_{0}(l) &=&\left( \frac{l-m}{l+m}\right) ^{\beta }, \\
\varphi _{0}(l) &=&\sqrt{\frac{\beta ^{2}-1}{2}}\ln \left[ \frac{l-m}{l+m}%
\right] .
\end{eqnarray}%
For a detailed analysis of this wormhole solution, see [16]. Eq.(15) can be
identified with Eq.(6) with $f=f_{0}$ provided we allow a further
identification $m\rightarrow im$. The variable $l\in (-\infty ,+\infty )$
gives two asymptotic flat solutions corresponding to $r=0$ and $r=\infty $.
The coordinate has a minimum value at the throat $l_{0}^{+}$ given by $%
l_{0}^{+}=r_{0}^{+}+\frac{m^{2}}{4r_{0}^{+}}=m\beta $ corresponding to $%
r=r_{0}^{+}$. Thus the minimum surface area at the throat is $4\pi
m^{2}\beta ^{2}$. For this solution, $n=4$, $p=m$, $q=\beta $, and using the
algorithm (9), the corresponding rotating EMS solution is%
\begin{equation}
f(l;m,\beta ,a)=\frac{8m\beta \delta f_{0}^{-1}}{a^{2}+4\delta ^{2}f_{0}^{-2}%
};\varphi (l)=\sqrt{\frac{\beta ^{2}-1}{2}}\ln \left[ \frac{l-m}{l+m}\right]
.
\end{equation}%
To achieve asymptotic flatness, that is, $f(l)\rightarrow 1$ as $%
l\rightarrow \pm \infty $, we note that $f_{0}(l)\rightarrow 1$ as $%
l\rightarrow \pm \infty $ . Therefore, we must fix%
\begin{equation}
\delta =\frac{2M\pm \sqrt{4M^{2}-a^{2}}}{2}.
\end{equation}%
In the above, we should retain only the positive sign before the square
root. The reason is the following: For $a=0$, the negative sign gives $%
\delta =0$ which implies $f=0$ which is meaningless. On the other hand, the
positive root gives $\delta =2M$ and $f=f_{0}$, as desired.

For the special case $\beta =1$, we have $\delta =\frac{2m\pm \sqrt{%
4m^{2}-a^{2}}}{2}$, and%
\begin{equation}
f(l;m,a)=\frac{8m\delta \left( \frac{l-m}{l+m}\right) ^{-1}}{a^{2}+4\delta
^{2}\left( \frac{l-m}{l+m}\right) ^{-2}}.
\end{equation}%
This is an asymptotically flat rotating solution without a scalar field, $%
\varphi =0$, but could it be interpreted as Kerr solution in some \textit{%
other} coordinates? The answer is not immediately evident. The coordinate
system in the metric (6) or (15) is itself rotating in the asymptotic region
($l\rightarrow \pm \infty $ ) where it is represented by%
\begin{equation}
ds^{2}=-\left( dt+a\cos \psi d\psi \right) ^{2}+dl^{2}+l^{2}\left( d\theta
^{2}+\sin ^{2}\theta d\psi ^{2}\right) .
\end{equation}%
We can retain terms to first order in $a$ and using Eq.(19) in Eq.(6), we
get the cross term as $\left( 1-\frac{2m}{l}\right) \times 2a\cos \theta
dtd\psi $. Now, we can subtract the rotational part of the coordinate system
which is given by $2a\cos \theta dtd\psi $. Then we are left with $\frac{4am%
}{l}\cos \theta $ which is not quite the Lense-Thirring term $\frac{4am}{l}%
\sin ^{2}\theta $ . However, such a direct subtraction is not a valid
procedure as the field equations are essentially nonlinear. Still, the
Eq.(19) is a formal solution of the field equations (7) and (8) under the
ansatz (6), though the question above does need a more detailed
investigation. This will be a separate task to be undertaken elsewhere.

To obtain the rotating BD solution, we follow the following steps: Note from
Eq.(3) that%
\begin{equation}
\sqrt{\frac{2\omega +3}{2}}\ln \phi =\varphi =\ln \left[ \frac{1-\frac{m}{2r}%
}{1+\frac{m}{2r}}\right] ^{\sqrt{2(\beta ^{2}-1)}}\Rightarrow \phi =\left[
\frac{1-\frac{m}{2r}}{1+\frac{m}{2r}}\right] ^{\sqrt{4(\beta
^{2}-1)/(2\omega +3)}}.
\end{equation}%
Now using the constraint from the BD field equations [9], viz.,%
\begin{equation}
4(\beta ^{2}-1)=-(2\omega +3)\frac{C^{2}}{\lambda ^{2}},
\end{equation}%
where $C,\lambda $ are two new arbitrary constants and $\omega $ is the
coupling parameter, we get%
\begin{equation}
\phi =\left[ \frac{1-\frac{m}{2r}}{1+\frac{m}{2r}}\right] ^{\frac{C}{\lambda
}}=\left[ \frac{l-m}{l+m}\right] ^{\frac{C}{2\lambda }}.
\end{equation}%
The Eq.(22) can be rephrased in the familiar form [9]:%
\begin{equation}
\lambda ^{2}=(C+1)^{2}-C(1-\frac{\omega C}{2}).
\end{equation}%
However, the wormhole condition $\beta ^{2}>1$ requires that the right hand
side of Eq.(22) be positive. This is possible if either $\omega <-\frac{3}{2}
$ or $\lambda $ be imaginary. Let us first consider $\omega <-\frac{3}{2}$
so that the exponents are real. Then, the final step consists in using the
relation $\widetilde{g}_{\mu \nu }=\phi ^{-1}g_{\mu \nu }$ together with
replacing $\beta $ in the exponents in the $g_{\mu \nu \text{ }}$by [7]%
\begin{equation}
\beta =\frac{1}{\lambda }\left( 1+\frac{C}{2}\right) .
\end{equation}%
This means, from Eq.(17), we have the BD rotating wormhole class I solution
for $\omega <-\frac{3}{2}$ as follows:%
\begin{eqnarray}
ds^{2} &=&-\widetilde{f}_{1}(l)dt^{2}+\widetilde{f}_{2}(l)\left[
dl^{2}+(l^{2}-m^{2})\left( d\theta ^{2}+\sin ^{2}\theta d\psi ^{2}\right) %
\right] , \\
\widetilde{f}_{1}(l) &\equiv &\widetilde{f}_{1}(l;m,C,\lambda
,a)=f(l;m,\beta ,a)\phi ^{-1}  \nonumber \\
&=&\frac{8m\delta \left[ \frac{1}{2\lambda }(C+2)\right] \left[ \frac{l-m}{%
l+m}\right] ^{-\frac{1}{\lambda }\left( 1+\frac{C}{2}\right) }}{%
a^{2}+4\delta ^{2}\left[ \frac{l-m}{l+m}\right] ^{-\frac{2}{\lambda }\left(
1+\frac{C}{2}\right) }}\times \left[ \frac{l-m}{l+m}\right] ^{-\frac{C}{%
2\lambda }}, \\
\widetilde{f}_{2}(l) &\equiv &f_{2}(l;m,C,\lambda ,a)=f^{-1}(l;m,\beta
,a)\phi ^{-1}  \nonumber \\
&=&\frac{a^{2}+4\delta ^{2}\left[ \frac{l-m}{l+m}\right] ^{-\frac{2}{\lambda
}\left( 1+\frac{C}{2}\right) }}{8m\delta \left[ \frac{1}{2\lambda }(C+2)%
\right] \left[ \frac{l-m}{l+m}\right] ^{-\frac{1}{\lambda }\left( 1+\frac{C}{%
2}\right) }}\times \left[ \frac{l-m}{l+m}\right] ^{-\frac{C}{2\lambda }}, \\
\phi (l) &=&\left[ \frac{l-m}{l+m}\right] ^{\frac{C}{2\lambda }}.
\end{eqnarray}%
It can be verified that the BD field equations again yield the expression
(23). Using the relation $l=r+\frac{m^{2}}{4r}$, it can be easily expressed
in the familiar ($t,r,\theta ,\psi $) coordinates with the value of $\delta $
given by Eq.(18) in which $\beta $ should have the value as in Eq.(25). For
instance, when $a=0$, we have $\delta =\frac{m}{\lambda }(C+2)$ and
identifying $\frac{m}{2}=B$, one retrieves the static BD metric in the
original notation:%
\begin{eqnarray}
ds^{2} &=&-\left( \frac{1-\frac{B}{r}}{1+\frac{B}{r}}\right) ^{\frac{2}{%
\lambda }}dt^{2}+\left( 1+\frac{B}{r}\right) ^{4}\left( \frac{1-\frac{B}{r}}{%
1+\frac{B}{r}}\right) ^{\frac{2(\lambda -C-1)}{\lambda }}\times   \nonumber
\\
&&[dr^{2}+r^{2}d\theta ^{2}+r^{2}\sin ^{2}\theta d\psi ^{2}] \\
\phi (r) &=&\left( \frac{1-\frac{B}{r}}{1+\frac{B}{r}}\right) ^{\frac{C}{%
\lambda }}.
\end{eqnarray}%
The condition for the above solution to represent a traversable wormhole is
[6]
\begin{equation}
(C+1)^{2}>\lambda ^{2}.
\end{equation}

For $\beta ^{2}>1$, and $\alpha =+1$, the negative kinetic term in the field
equations (5) shows that the energy density is negative violating the Weak
Energy Condition (WEC) so that the Eqs.(6), (15) and (17) provide a class of
rotating EMS wormhole solution. This solution is then mapped into the BD
regime given by the Eqs.(24)-(27) for the range of coupling values $\omega <-%
\frac{3}{2}$. Same classes of solutions will be obtained by alternative
calculations with $\beta ^{2}>1$ and $\alpha =-1$ (imaginary scalar charge).
One could also consider Eq.(11) with the values $\alpha =-1$ (positive
kinetic term) and $\beta ^{2}<1$. Then the above procedure would produce a
rotating naked singularity in EMS and BD theory. The above calculations
represent the basic scheme to be followed in other classes of EMS or BD
solutions.

\begin{center}
\textbf{IV. BD Class II rotating solution}
\end{center}

Next consider the imaginary value of $\lambda $ in Eq.(21) with $\omega >-%
\frac{3}{2}$. Let us take $\lambda \equiv -i\Lambda $. We prefer to start
with the BD solutions (30), (31) and then obtain therefrom the EMS seed
solution. The reason is to show that BD class I and II solutions are not
independent. Thus, to make Eq.(30) real, it is now necessary to take $%
B\equiv ib$ and use the identity%
\begin{equation}
\arctan (x)=\frac{i}{2}\ln \left( \frac{1-ix}{1+ix}\right) .
\end{equation}%
where $x$ is real. Using further $r\rightarrow \frac{1}{r}$, we can finally
rewrite (30) and (31) as%
\begin{eqnarray}
ds^{2} &=&-Exp\left[ 2\alpha _{0}+\frac{4}{\Lambda }\arctan \left( \frac{r}{b%
}\right) \right] dt^{2}+  \nonumber \\
&&Exp\left[ 2\beta _{0}-\frac{4(C+1)}{\Lambda }\arctan \left( \frac{r}{b}%
\right) -2\ln \left( \frac{r^{2}}{r^{2}+b^{2}}\right) \right] \times
\nonumber \\
&&[dr^{2}+r^{2}d\theta ^{2}+r^{2}\sin ^{2}\theta d\psi ^{2}], \\
\phi (r) &=&Exp\left[ \frac{2C}{\Lambda }\arctan \left( \frac{r}{b}\right) %
\right] ,
\end{eqnarray}%
where $\alpha _{0}$ and $\beta _{0}$ are two adjustable constants to be
determined by the asymptotic flatness. Defining $\gamma =\frac{1}{\Lambda }%
\left( 1+\frac{C}{2}\right) $, using the relation $\widetilde{g}_{\mu \nu
}=\phi ^{-1}g_{\mu \nu }$ and the redefinition $\phi \rightarrow \varphi $
via Eq.(3), we obtain the exact EMS class II solution:%
\begin{eqnarray}
ds^{2} &=&-Exp\left[ 2\alpha _{0}+4\gamma \arctan \left( \frac{r}{b}\right) %
\right] dt^{2}+  \nonumber \\
&&\left( 1+\frac{b^{2}}{r^{2}}\right) ^{2}Exp\left[ 2\beta _{0}-4\gamma
\arctan \left( \frac{r}{b}\right) \right] \times   \nonumber \\
&&[dr^{2}+r^{2}d\theta ^{2}+r^{2}\sin ^{2}\theta d\psi ^{2}], \\
\varphi (r) &=&\sqrt{8(1+\gamma ^{2})}\arctan \left( \frac{r}{b}\right) .
\end{eqnarray}

Asymptotic flatness requires that $\alpha _{0}=-\pi \gamma $, $\beta
_{0}=\pi \gamma $. This also represents a traversable wormhole with the
throat appearing at $r_{0}^{+}=b\left[ \gamma +\sqrt{1+\gamma ^{2}}\right] $%
. The metric functions expand exactly as in Eq.(12) and the mass of the
wormhole is $M=2b\gamma $. The solution can be rephrased using $l=r-\frac{%
b^{2}}{r}$as the seed solution like in Eqs.(15), (16):
\begin{eqnarray}
ds^{2} &=&-f_{0}(l)dt^{2}+\frac{1}{f_{0}(l)}\left[ dl^{2}+(l^{2}+4b^{2})%
\left( d\theta ^{2}+\sin ^{2}\theta d\psi ^{2}\right) \right] , \\
f_{0}(l) &=&Exp\left[ 4\gamma \left\{ -\frac{\pi }{2}+\arctan \left( \frac{l+%
\sqrt{l^{2}+4b^{2}}}{2b}\right) \right\} \right] , \\
\varphi _{0}(l) &=&\sqrt{8(1+\gamma ^{2})}\arctan \left( \frac{l+\sqrt{%
l^{2}+4b^{2}}}{2b}\right) ,
\end{eqnarray}%
where $b,\gamma $ are the constant parameters of the solution. The minimum
surface area in the $l$ coordinate is $4\pi (2b\gamma )^{2}$. The field
Eqs.(7) and (8) are satisfied by the solution (39), (40) and we have also
seen how the class II solution can be derived from the class I solution. For
this solution $n=\frac{1}{4},p=2b,q=4\gamma $ and using the algorithm (9),
the corresponding rotating EMS solution is Eq.(6) with $l_{0}=2b$ and
\begin{equation}
f(l;b,\gamma ,a)=\frac{16b\gamma \delta f_{0}^{-1}}{4a^{2}+\delta
^{2}f_{0}^{-2}};\varphi (l)=\sqrt{8(1+\gamma ^{2})}\arctan \left( \frac{l+%
\sqrt{l^{2}+4b^{2}}}{2b}\right) .
\end{equation}

To achieve asymptotic flatness, that is, $f(l)\rightarrow 1$ as $%
l\rightarrow \pm \infty $, we note that $f_{0}(l)\rightarrow 1$ as $%
l\rightarrow \pm \infty $. This information immediately gives us%
\begin{equation}
\delta =2\left[ 2M+\sqrt{4M^{2}-a^{2}}\right] .
\end{equation}%
This case represents wormhole spacetime in which exotic matter is spread
throughout the spacetime though with asymptotically decaying density. A more
realistic situation is to confine the rotating matter within fixed limits $%
l=l_{B\pm }$ while allowing for a flat spacetime beyond those limits. Then,
we need to have a matching on two sides at $l=l_{B\pm }$ and obtain values $%
\delta _{\pm }$, one for each side. We first match the static case for $l\in
\lbrack l_{B+},l_{B-}]$ by introducing a constant $\varepsilon $ in Eq.(39)
as%
\begin{eqnarray}
f_{0}(l) &=&Exp\left[ 4\gamma \left\{ \varepsilon +\arctan \left( \frac{l+%
\sqrt{l^{2}+4b^{2}}}{2b}\right) \right\} \right] ,  \nonumber \\
\varphi _{0}(l) &=&\sqrt{8(1+\gamma ^{2})}\arctan \left( \frac{l+\sqrt{%
l^{2}+4b^{2}}}{2b}\right) .
\end{eqnarray}%
The constant can be chosen as follows: $\varepsilon =+\frac{\pi }{2}$ for $%
l\in \lbrack l_{B+},\infty )$ and $\varepsilon =-\frac{\pi }{2}$ for $l\in
\lbrack l_{B-},-\infty )$. Thus at the upper boundary $l=l_{B+}$,%
\begin{eqnarray}
f_{0+} &=&f_{0}(l_{B+})=Exp\left[ 4\gamma \left\{ +\frac{\pi }{2}+\arctan
\left( \frac{l_{B+}+\sqrt{l_{B+}^{2}+4b^{2}}}{2b}\right) \right\} \right] ,
\nonumber \\
\varphi _{0+} &=&\varphi _{0}(l_{B+})=\sqrt{8(1+\gamma ^{2})}\arctan \left(
\frac{l_{B+}+\sqrt{l_{B+}^{2}+4b^{2}}}{2b}\right) ,
\end{eqnarray}%
and at the lower boundary $l=l_{B-}$,%
\begin{eqnarray}
f_{0-} &=&f_{0}(l_{B-})=Exp\left[ 4\gamma \left\{ -\frac{\pi }{2}+\arctan
\left( \frac{l_{B-}+\sqrt{l_{B-}^{2}+4b^{2}}}{2b}\right) \right\} \right] ,
\nonumber \\
\varphi _{0-} &=&\varphi _{0}(l_{B-})=\sqrt{8(1+\gamma ^{2})}\arctan \left(
\frac{l_{B-}+\sqrt{l_{B-}^{2}+4b^{2}}}{2b}\right) .
\end{eqnarray}%
Now, to match the rotating EMS solution to those boundaries, we need:%
\begin{equation}
f_{\pm }\equiv f(l_{B\pm },b,\gamma ,a,\delta _{\pm })=f_{0\pm },
\end{equation}%
in which case the constants are determined by%
\begin{equation}
\delta _{\pm }=2\left[ 2M+\sqrt{4M^{2}-a^{2}f_{0\pm }^{2}}\right] .
\end{equation}%
These reduce to the same expression for $\delta $ as in Eq.(42) above if $%
l_{B\pm }\rightarrow \infty $. In the intermediate region , we have%
\begin{equation}
f=f(l;b,\gamma ,a,\varepsilon ,\delta )
\end{equation}%
where $\delta _{-}\leq \delta \leq \delta _{+}$, $-\frac{\pi }{2}\leq
\varepsilon \leq \frac{\pi }{2}$.

To obtain the rotating BD wormhole class II solution, one has to replace $%
\gamma $ by $\gamma =\frac{1}{\Lambda }\left( 1+\frac{C}{2}\right) $ in the
solution (38) and use $\Lambda ^{2}=C(1-\frac{\omega C}{2})-(C+1)^{2}$. The
last relation follows directly from Eq.(22) when $\lambda \equiv -i\Lambda $%
. The complete formal solution can be obtained by using Eqs.(34), (38) and
the first equation in (41) as%
\begin{eqnarray}
ds^{2} &=&-\widetilde{f_{1}}(l)\left( dt+a\cos \psi d\psi \right) ^{2}+%
\widetilde{f_{2}}(l)\left[ dl^{2}+(l^{2}+4b^{2})\left( d\theta ^{2}+\sin
^{2}\theta d\psi ^{2}\right) \right] ,  \nonumber \\
\widetilde{f_{1}}(l) &\equiv &\widetilde{f}_{1}(l;b,C,\Lambda
,a)=f(l;b,\gamma ,a)\phi ^{-1}  \nonumber \\
&=&\frac{2b\delta \left[ \frac{1}{\Lambda }(C+2)\right] Exp\left[ -\frac{%
2\pi }{\Lambda }-\frac{4}{\Lambda }\arctan \left( \frac{l+\sqrt{l^{2}+4b^{2}}%
}{2b}\right) \right] }{a^{2}+\frac{1}{4}\delta ^{2}Exp\left[ -\frac{4\pi }{%
\Lambda }-\frac{8}{\Lambda }\arctan \left( \frac{l+\sqrt{l^{2}+4b^{2}}}{2b}%
\right) \right] }\times   \nonumber \\
&&Exp\left[ -\frac{2C}{\Lambda }\arctan \left( \frac{l+\sqrt{l^{2}+4b^{2}}}{%
2b}\right) \right] , \\
\widetilde{f_{2}}(l) &\equiv &\widetilde{f}_{2}(l;b,C,\Lambda
,a)=f^{-1}(l;b,\gamma ,a)\phi ^{-1}, \\
\phi (l) &=&Exp\left[ \frac{2C}{\Lambda }\arctan \left( \frac{l+\sqrt{%
l^{2}+4b^{2}}}{2b}\right) \right] ,
\end{eqnarray}%
where $\delta $ is given by Eq.(42) to ensure asymptotic flatness.

\begin{center}
\textbf{V. BD Class IV rotating solution}
\end{center}

We start with the static EMS class IV solution as the seed solution [10]:%
\begin{eqnarray}
ds^{2} &=&-f_{0}(l)dt^{2}+\frac{1}{f_{0}(l)}\left[ dl^{2}+l^{2}\left(
d\theta ^{2}+\sin ^{2}\theta d\psi ^{2}\right) \right] , \\
f_{0}(l) &=&Exp\left[ -\frac{\gamma }{bl}\right] ,\phi _{0}(l)=-\frac{\gamma
}{\sqrt{2}bl},M=\frac{\gamma }{2b}.
\end{eqnarray}%
For this solution, $n=4$, $p=\gamma $, $q=\frac{1}{2b}$. This solution
represents an asymptotically flat traversable wormhole with the minimum
surface area $4\pi \left( \frac{\gamma }{2b}\right) ^{2}$. Thus the rotating
EMS solution is:%
\begin{equation}
f(l;b,\gamma ,a)=\frac{4\gamma \delta e^{\frac{\gamma }{bl}}}{b\left(
a^{2}+4\delta ^{2}e^{\frac{2\gamma }{bl}}\right) };\varphi (l)=-\frac{\gamma
}{\sqrt{2}bl}.
\end{equation}%
Asymptotic flatness fixes $\delta =\frac{2M+\sqrt{4M^{2}-a^{2}}}{2}$ and
redefine the EMS constants $\gamma ,b$ as%
\begin{equation}
\frac{\gamma }{b}=\frac{C+2}{B}
\end{equation}%
and then use the BD constraint [$\lambda =0$ in Eq.(24)]:%
\begin{equation}
C=\frac{-1\pm \sqrt{-2\omega -3}}{\omega +2},
\end{equation}%
such that, following similar arguments surrounding Eqs.(21)-(25), for $%
\omega <-\frac{3}{2}$, we have real values for $C$ and
\begin{equation}
\phi (l)=Exp\left[ -\frac{C}{Bl}\right] .
\end{equation}%
The rotating BD IV metric is:%
\begin{eqnarray}
\widetilde{f_{1}}(l) &\equiv &\widetilde{f}_{1}(l;B,C,a)=f(l;b,\gamma
,a)\phi ^{-1}  \nonumber \\
&=&\frac{\frac{4\delta }{B}(C+2)Exp\left[ \frac{C+2}{Bl}\right] }{%
a^{2}+4\delta ^{2}Exp\left[ \frac{2(C+2)}{Bl}\right] }\times Exp\left[ \frac{%
C}{Bl}\right] , \\
\widetilde{f_{2}}(l) &\equiv &\widetilde{f}_{2}(l;B,C,a)=f^{-1}(l;b,\gamma
,a)\phi ^{-1}
\end{eqnarray}

There is not much to say about the class III solution. The EMS solution can
be obtained from the class IV EMS solution under the same constraint (56).
All that one has to do is invert $l\rightarrow \frac{1}{l}$ so that
\begin{eqnarray}
ds^{2} &=&-f_{0}(l)dt^{2}+g_{0}(l)\left[ dl^{2}+l^{2}\left( d\theta
^{2}+\sin ^{2}\theta d\psi ^{2}\right) \right] , \\
f_{0}(l) &=&Exp\left[ -\frac{\gamma l}{b}\right] ,g_{0}(l)=\left( \frac{l}{b}%
\right) ^{-4}Exp\left[ \frac{\gamma l}{b}\right] \phi _{0}(l)=\frac{\gamma l%
}{\sqrt{2}b}.
\end{eqnarray}%
The above solution is not asymptotically flat though it is flat at $l=0$.
Therefore, it does not meet the requirement of the asymptotic
\textquotedblleft flaring out" condition for traversable wormholes. Hence,
we do not discuss the solution further including its rotating BD version.

\begin{center}
\textbf{VI. Rotating wormholes in string theory}
\end{center}

Formal rotating solutions in the string theory can also be obtained via the
conformal transformation%
\begin{equation}
h_{1}(l)=\phi ^{-1}f(l)=e^{-2\varphi }f(l),h_{2}(l)=\phi
^{-1}f^{-1}(l)=e^{-2\varphi }f^{-1}(l),
\end{equation}%
where $\phi (l)$ and $f(l)$ are the BD rotating scalar and EMS rotating
metric solutions respectively. The complete solution is%
\begin{eqnarray}
ds^{2} &=&-h_{1}(l)\left( dt+a\cos \psi d\psi \right) ^{2}+h_{2}(l)\left[
dl^{2}+(l^{2}+l_{0}^{2})\left( d\theta ^{2}+\sin ^{2}\theta d\psi
^{2}\right) \right] , \\
\widetilde{\Phi } &=&-\frac{1}{\sqrt{2}}\varphi .
\end{eqnarray}%
The values of $l_{0}^{2}$ are $-m^{2}$, $4b^{2}$ and $0$ for classes I, II
and IV string solutions respectively. The corresponding values for $f(l)$
and $\varphi $ can be taken from the Eq.(17) above. We display only class I
rotating string solution here:%
\begin{eqnarray}
h_{1}(l) &=&=\frac{8m\beta \delta \left( \frac{l-m}{l+m}\right) ^{-(\beta +%
\sqrt{\beta ^{2}-1})}}{a^{2}+4\delta ^{2}\left( \frac{l-m}{l+m}\right)
^{-2\beta }};h_{2}(l)==\frac{a^{2}+4\delta ^{2}\left( \frac{l-m}{l+m}\right)
^{-2\beta }}{8m\beta \delta \left( \frac{l-m}{l+m}\right) ^{-\beta +\sqrt{%
\beta ^{2}-1}}}, \\
\widetilde{\Phi }(l) &=&-\frac{1}{2}\sqrt{\beta ^{2}-1}\ln \left[ \frac{l-m}{%
l+m}\right] ,
\end{eqnarray}%
where $\delta $ is given by Eq.(18) with $M=m\beta $. Consider the static
case, $a=0$. Using the identities
\begin{equation}
l^{2}-m^{2}\equiv r^{2}\left( 1-\frac{m^{2}}{4r^{2}}\right) ^{2};\frac{l-m}{%
l+m}\equiv \left[ \frac{1-\frac{m}{2r}}{1+\frac{m}{2r}}\right] ^{2},
\end{equation}%
we may retrieve the solution for static stringy wormholes derived in
Ref.[12]. Identifying the exponents as $E=\pm \beta $ and $F=\pm \sqrt{\beta
^{2}-1}$, we have $E^{2}-F^{2}=1$. This is actually the constraint provided
by the string field equations in the wormhole case. In the case of naked
singularity,  $E=\pm \beta $ and $F=\pm \sqrt{1-\beta ^{2}}$ and in this
case, the string field equations give $E^{2}+F^{2}=1$, as can be verified
from the works in Ref.[12].

Other classes of solutions can be derived likewise. The rotating solution
for naked singularity can be obtained simply by choosing $\beta ^{2}<1$
without any extra effort.

\begin{center}
\textbf{VII. Conclusions }
\end{center}

Asymptotically flat rotating solutions are rather rare in the literature, be
it of a wormhole or naked singularity. The present article has provided
\textit{formal} asymptotically flat rotating solutions in the EMS and BD
theories with extensions to string theory. There is however a caveat here.
Do the solutions really represent a physically rotating configuration? We
can only give a partial answer. Certainly, the solutions are not the result
of just writing down a known static metric in rotating coordinate systems,
but more -- they \textit{do} contain information about the rotation of the
physical configurations in question.

The solutions represent mathematically interesting features of EMS and BD
field equations. The string solutions are just the BD solutions with $\omega
=-1$. As we saw, the solutions admit two arbitrary parameters $a$ and $%
\delta $. The quantity $a$ has been interpreted by Matos and N\'{u}\~{n}ez
[8] as a rotation parameter of the gravity field. However, we think that
their ansatz, that we have used here, represents a \textit{nonlinear mixture}
of the rotation of coordinate frame and the rotation of the gravity field
due to wormhole or naked singularity. The other parameter $\delta $ is fixed
either by asymptotic flatness or by the desired matching conditions. Further
investigations into the nature of solutions with a view to separating the
real rotational effects from the fictitious effects arising out of the
coordinate rotation might be rewarding.

\textbf{Acknowledgments:}

One of the authors (KKN) wishes to acknowledge financial assistance from
TWAS, Italy under its associateship program. Thanks are due to Guzel
Kutdusova for helpful comments.

\textbf{References}

[1] M.S. Morris and K.S. Thorne, Am. J. Phys. \textbf{56}, 395 (1988); M.S.
Morris, K.S. Thorne and U. Yurtsever, Phys. Rev. Lett. \textbf{61}, 1446
(1988).

[2] A. Einstein and N. Rosen, Phys. Rev. \textbf{48}, 73 (1935).

[3] D. Hochberg, Phys. Lett. B \textbf{251}, 349 (1990).

[4] M. Visser, \textit{Lorentzian Wormholes- From Einstein to Hawking}
(A.I.P., New York, 1995).

[5] A.G. Agnese and M. La Camera, Phys. Rev. D \textbf{51}, 2011 (1995);
K.K. Nandi, A. Islam and J. Evans, Phys. Rev. D \textbf{55}, 2497 (1997).

[6] C. Barcel\'{o} and M. Visser, Class. Quant. Grav. \textbf{17}, 3843
(2000); B. McInnes, J. High Energy Phys. \textbf{12}, 053 (1002); G.
Klinkhammer, Phys. Rev. D \textbf{43}, 2512 (1991); L. Ford and T.A. Roman,
Phys. Rev. D \textbf{46}, 1328 (1992); \textit{ibid}, \textbf{48}, 776
(1993); L.A. Wu, H.J. Kimble, J.L. Hall and H. Wu, Phys. Rev. Lett. \textbf{%
57}, 2520 (1986).

[7] K.K. Nandi, B. Bhattacharjee, S.M.K. Alam and J. Evans, Phys. Rev. D
\textbf{57}, 823 (1998); L. Anchordoqui, S.P. Bergliaffa and D.F. Torres,
Phys. Rev. D \textbf{55}, 5226 (1997).

[8] T. Matos and D. N\'{u}\~{n}ez, gr-qc/0508117 and references therein; T.
Matos, Gen. Rel. Grav. \textbf{19}, 481 (1987).

[9] C.H. Brans, Phys. Rev. \textbf{125}, 2194 (1962).

[10] A. Bhadra and K.K. Nandi, Mod. Phys. Lett. A \textbf{16}, 2079 (2001).

[11] A. Bhadra and K. Sarkar, gr-qc/0505141.

[12] S. Kar, Class. Quant. Grav. \textbf{16}, 101 (1999)

[13] Y.M. Cho, Phys. Rev. Lett. \textbf{68}, 3133 (1992).

[14] H.A. Buchdahl, Phys. Rev. \textbf{115}, 1325 (1959).

[15] C. Armend\'{a}riz-P\'{\i}con, Phys. Rev. D \textbf{65}, 104010 (2002).

[16] K.K. Nandi, Y.-Z. Zhang, Phys. Rev. D \textbf{70}, 044040 (2004); K.K.
Nandi, Y.-Z. Zhang and K.B. Vijay Kumar, Phys. Rev. D \textbf{70}, 127503
(2004).

\end{document}